\documentclass[preprint]{raa}
\usepackage{graphicx,times}             
\usepackage{natbib}
\usepackage{amssymb,amsmath}
\usepackage{epstopdf}
\bibpunct{(}{)}{;}{a}{}{,}

\usepackage{threeparttable}
\usepackage{float}
\usepackage[a4paper=true,pagebackref=true]{hyperref}
\hypersetup{colorlinks = true, linkcolor = green, anchorcolor = red, citecolor = blue, filecolor = red, pagecolor = red, urlcolor = red}

\usepackage{color}
\usepackage{soul}

\begin{document}

   \title{The abundance of massive compact galaxies at $1.0 < z < 3.0$
    in 3D-\textit{HST}/CANDELS\,$^*$
\footnotetext{$*$ Supported by the National Natural Science Foundation of China.}
}

   \volnopage{Vol.0 (20xx) No.0, 000--000}      
   \setcounter{page}{1}          

   \author{Shi-Ying Lu
      \inst{1}
   \and Yi-Zhou Gu
      \inst{1}
   \and Guan-Wen Fang
      \inst{2}
   \and Qi-Rong Yuan
      \inst{1}
   }

   \institute{Department of Physics and Institute of Theoretical Physics
             Nanjing Normal University, Nanjing 210023,China;
             {\it yuanqirong@njnu.edu.cn}\\
        \and
             Institute for Astronomy and History of Science and Technology,
             Dali University, Dali 671003,China;
             {\it wen@mail.ustc.edu.cn}\\
\vs\no
   {\small Received~~20xx month day; accepted~~20xx~~month day}}

\abstract{Based on a large sample of massive ($M_{\ast}\geqslant10^{10} M_{\odot}$) compact galaxies at $1.0 < z < 3.0$ in five 3D-\textit{HST}/CANDELS fields, we quantify the fractional abundance and comoving number density of massive compact galaxies as a function of redshift.
The samples of compact quiescent galaxies (cQGs) and compact star-forming galaxies (cSFGs) are constructed by various selection criteria of compact galaxies in literatures, and the effect of compactness definition on abundance estimate is proved to be remarkable, particularly for the cQGs and cSFGs at high redshifts.
Regardless of the compactness criteria adopted, their overall redshift evolutions of fractional abundance and number density are found to be rather similar.
Large samples of the cQGs exhibit a sustaining increase in number density from $z \sim 3$ to 2 and a plateau at $1<z<2$. For massive cSFGs, a plateau in the number density at $2<z<3$ can be found, as well as a continuous drop from $z \sim 2$ to 1. The evolutionary trends of the cQG and cSFG abundances support the scenario that the cSFGs at $z \gtrsim 2$ may have been rapidly quenched into quiescent phase via violent dissipational processes such as major merger and disk instabilities.
Rarity of the cSFGs at lower redshifts  ($z < 1$) can be interpreted by the decrease of gas reservoirs in dark matter halos and the consequent low efficiency of gas-rich dissipation.
}
\vfill
   \authorrunning{S.-Y. Lu, G.-W. Fang, Y.-Z. GU, \& Q.-R. Yuan }  
   \titlerunning{The abundance of massive compact galaxies at $1.0 < z < 3.0$ in 3D-\textit{HST}/CANDELS} 

   \maketitle

%
%
\newpage
\section{Introduction}           
\label{sect:intro}

It has been widely appreciated that there is a bimodality of galaxy populations, i.e., blue star-forming galaxies (SFGs) vs. red quiescent galaxies (QGs), since the universe was only $\sim$2.5 Gyr old (\citealt{Strateva+2001,Kauffmann+2003a, Kauffmann+2003b, Baldry+2004,Blanton+Moustakas+2009,Brammer+2011,Whitaker+2011,Whitaker+2012,Huertas-Company+2015}). It is believed that there should be an evolutionary connection between the two populations. A picture of star formation quenching has been proposed that the SFGs would truncate their star formation activities and transform into a quiescent status (\citealt{Blanton+03,Brinchmann+04, Kauffmann+04, Faber+07, Peng+10, Fang+2012,Barro+2013, Barro+2014,Gu+2018a}).
A lot of large surveys (such as the SDSS, NMBS, UltraVISTA,  zFOURGE, and CANDELS) have provided the probability to study the physical processes and mechanisms relevant to star formation quenching over a wide span of cosmic time.

Observational link between quenching and structure properties has caught more and more attention. In general, the SFGs are found to have an extended structure, with larger non circularized effective radii ($r_{\rm e}$) than the QGs (e.g, \citealt{Williams+2009,Fang+2012, van+der+Wel+2012,Whitaker+2012,Cassata+2013,van+der+Wel+2014,Huertas-Company+2015}).
The quiescent galaxies in the early epoch are three to five times more compact than their local counterparts (\citealt{Newman+2010,Bruce+2012, Ryan+2012, Cassata+2013}).
Moveover, \cite{van+der+Wel+2014} report that early-type galaxies (ETGs) at the fixed stellar mass follow a faster size evolution, $ r_{\rm e} \varpropto (1+z)^{-1.48} $, while late-type galaxies (LTGs) follow a slower evolution of size, $ r_{\rm e} \varpropto (1+z)^{-0.75}$.
Compact quiescent galaxies (cQGs, also called ``red nuggets'') are found to be ubiquitous at $z\sim 2$ (\citealt{Damjanov+2009}). A similar population of compact star-forming galaxies (cSFGs, also called``blue nugget'') is confirmed to be presence at high redshifts (\citealt{Barro+2013, Barro+2014, Fang+2015, van+Dokkum+2015}). However, in the local universe, massive compact galaxies are quite rare with the number density in the order of magnitude $10^{-6} \ \rm Mpc^{-3}$ (\citealt{Trujillo+2009, Taylor+10, Trujillo+2014, Graham+2015, Saulder+2015, Buitrago+18}) but prefer to be found in galaxy clusters (\citealt{Valentinuzzi+10, Poggianti+13a, Poggianti+13b, Peralta+16}).

It remains unsolved why the abundances of compact galaxies are  discrepant at different redshifts and how these massive compact galaxies form and evolve. Some mechanisms are proposed to explain the formation and evolution of these compact galaxies.
Star-forming galaxies with extended structures (called extended SFGs, eSFGs for short) are believed to be the progenitors of massive compact galaxies (\citealt{Barro+2013, Barro+2014, Fang+2015, van+Dokkum+2015}). It suggests that the cSFGs are formed from the eSFGs by shrinking their sizes via the violent gas-rich dissipational processes (\citealt{Dekel+13, Dekel+14, Zolotov+2015}).
On account of the high luminosities of star formation or active galactic nucleus (AGN) activities triggered by gas-rich dissipational processes, the cSFGs would consume their cools gas rapidly, and evolve to the cQGs soon (\citealt{Barro+2013,Fang+2015, Tadaki+15}). Furthermore, these cQGs could evolve to the local massive quiescent galaxies or extend quiescent galaxies (eQGs) through minor mergers later (\citealt{Hopkins+10,delaRosa+16}). All in all, the majority of these massive compact galaxies at $z \sim 2$ end up in the central dense cores of the local galaxies (\citealt{van+Dokkum+2014, Belli+2014a}).

Although the compact galaxy population covering a wide range of redshift has been studied by many investigators (e.g, $z\lesssim 1.0$: \citealt{Trujillo+2009, Trujillo+2014,Saulder+2015,Zahid+2015,Charbonnier+2017}; $z\gtrsim 1.0$: \citealt{Barro+2014,Cassata+2013,Fang+2015,van+der+Wel+2014,van+Dokkum+2015}), the statistical results of massive compact galaxies are rather diverse, which is mainly due to different observational data and different strategies in selection of compact galaxies. For example, \cite{Charbonnier+2017} and \cite{Damjanov+2018inpreparation} applied same criteria of compact galaxies to the CFHT Stripe 82 (CS82) survey and the Subaru Hyper Suprime-Cam (HSC) high-resolution imaging survey, respectively. Their cosmic evolution of cQG number densities since $z \sim 0.4$ is different from each other.
Even with the same data, the statistics of massive compact galaxies (i.e., cQGs and cSFGs) will be severely biased when we adopt different thresholds of stellar mass, effective radius, and compactness in sampling. For instance, abundance of the cSFGs in the CANDELS fields at higher redshifts ($z\gtrsim 1.0$) has been estimated by \cite{Barro+2014} and \cite{Fang+2015} with different compactness criteria, and their results are different.

To untangle the effect of different compactness criteria, it is necessary to make a comprehensive comparison for the samples of cSFGs and cQGs at higher redshifts that are selected with different criteria.
In this paper we will form a large sample of massive $(M_{\ast} \geqslant 10^{10} M_{\odot})$ galaxies at $1.0 < z < 3.0$ in the five deep fields of the 3D-\textsl{HST}/ \textrm{CANDELS} programs (\citealt{Grogin+2011, Koekemoer+2011, Skelton+2014}). All massive galaxies are separated into quiescent and star-forming populations using the rest-frame \textrm{UVJ} diagram (\citealt{Williams+2009}).
Then, eight different criteria of compact galaxies in the literatures (\citealt{Carollo+2013,Quilis+Trujillo+2013,Barro+2014,van+der+Wel+2014,Fang+2015,van+Dokkum+2015}) will be adopted to construct the samples of cQGs and cSFGs.
For these various samples of cSFGs and cQGs, their fractional abundances and number densities can be computed, as a function of redshift.
A detailed comparison between these results can tell us how the different criteria affect the conclusions about fractional abundance and number density of cQGs and cSFGs.

The paper is organized as follows. We give an overview of the 3D-\textsl{HST}/CANDELS data set and a description of our sample selection in Section \ref{sect:data and sample selection}, including various criteria of compact galaxies.
In Section \ref{sect:The Abundance of massive compact galaxies}, we present the evolution of the fraction and number density of massive compact galaxies, and further discuss the evolutionary connection between cSFGs and cQGs.
Finally, we give a summary in Section \ref{sect:summary}. Throughout the paper, we assume the cosmology model with $\rm \Omega_{M} = 0.3$, $\rm \Omega_{\Lambda} = 0.7$, and $\rm H_{0} = 70\, km\;s^{-1} Mpc^{-1} $.

\section{Data and Sample Selection}
\label{sect:data and sample selection}

\subsection{Data Description}
\label{sect2.1:Data Description }
On the basis of the high-quality \textrm{WFC3} and \textrm{ACS} spectroscopy and multi-wavelength photometry in the five 3D-\textit{HST}/\textrm{CANDELS} fields (i.e., \textrm{AEGIS}, \textrm{COSMOS}, \textrm{GOODS-N}, \textrm{GOODS-S}, and \textrm{UDS}) (\citealt{Grogin+2011, Koekemoer+2011, Skelton+2014}), we set about selecting a large sample of massive galaxies.
The data base is from the
\textrm{CANDELS} and 3D-\textit{HST} Treasury programs include the \textrm{WFC3} \textit{F}125\textit{W}, \textit{F}140\textit{W}, \textit{F}160\textit{W} images (\citealt{Skelton+2014}), which are observed with many other space- and ground-based telescopes. Total area of the five fragmented deep fields is $\sim$900 $ \textrm{arcmin}^2$, which can mitigate the influence of cosmic variance  to a certain extent.

The photometric redshifts $(z_{\textrm{phot}})$ and the rest-frame \textrm{UVJ} colors are derived by \cite{Skelton+2014} with the \textrm{EAZY} code (\citealt{Brammer+2008}). The derived photometric redshifts for the five CANDELS fields have higher precision, and their normalized median absolute deviations ($\sigma_{\textrm{NMAD}}$), defined as $\sigma_{\textrm{NMAD}}= 1.48 \times \textrm{median} [| \vartriangle z - {\rm median}(\vartriangle z)|/ ( 1 + z)]$, are within a range from 0.007 (\textrm{COSMOS}) to 0.026 (\textrm{GOODS-N}) (\citealt{Skelton+2014}).
In this paper, we preferentially adopt the spectroscopic redshifts ($z_{\rm spec}$) if available. The stellar masses are derived by \cite{Skelton+2014}, who fit the spectra energy distribution (SED) using the \textrm{FAST} code (\citealt{Kriek+2009}) based on the \cite{Bruzual+Charlot+2003} stellar population synthesis (\textrm{SPS}) models with the \cite{Chabrier+2003} initial mass function (\textrm{IMF}) and solar metallicity. Additionally, \cite{van+der+Wel+2012} have estimated the non circularized effective radius $r_{\rm e}$ and axis ratio $q$ by using the \textrm{GALFIT}  code (\citealt{Peng+02}). We adopt the \textrm{GALFIT} results of  J band (\textit{F}125\textit{W}) images for the galaxies at $1.0 < z \leqslant 1.8$, and  H band (\textit{F}160\textit{W}) results for the galaxies at $1.8 < z < 3.0$, for ensuring  the structural feature observed with the same optical band in the rest frame. The axis ratio $q$ can be taken to calculate the non circularized effective radius, a key parameter in some definitions of compactness (see Section \ref{sect2.3:Compactness critera}).

\subsection{Sample of Massive QGs and SFGs }
\label{sect2.2:Sample of Massive QGs and SFGs}
\begin{figure}
   \centering
   \includegraphics[width=11cm, angle=0]{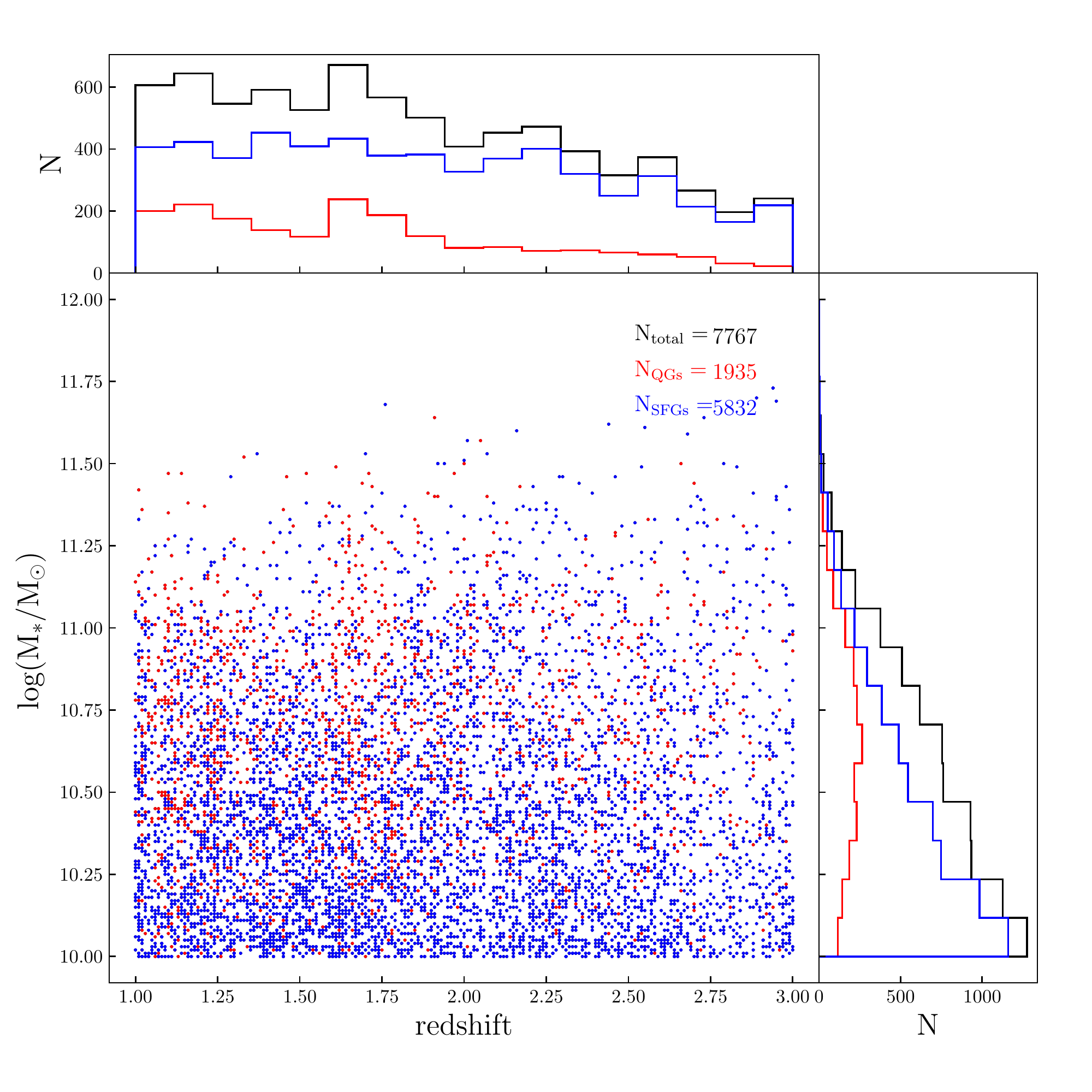}
   \caption{Distribution of the massive  galaxies in the five CANDELS fields in the stellar mass - redshift plane. The total massive galaxies (black dots/curves) are separated by \textrm{UVJ} diagram (see Section \ref{sect2.2:Sample of Massive QGs and SFGs} and Figure \ref{Fig2:UVJ}) into quiescent galaxies (red dots/curves) and star-forming ones (blue dots/curves). The corresponding numbers are shown at the top-right corner. }
   \label{Fig1:zlmass}
   \end{figure}

Firstly, based on the multiwavelength photometric data in five 3D-\textit{HST}/\textrm{CANDELS} fields, we select a large sample of 7767 massive galaxies ($M_{\ast} \geqslant 10^{10} M_{\odot}$) with good photometry qualities (i.e., \texttt{use}\_\verb"phot"=\texttt{1}) and good morphological fits with \textrm{GALFIT} (i.e., \textrm{GALFIT} flag $=0$ or$1$) at  $1.0 < z < 3.0$ to ensure high sample completeness and robust structural measurements.
The completeness above the mass threshold is around $\sim 90\%$ up to the highest redshift (\citealt{Grogin+2011,Wuyts+11,Newman+12,Barro+2013,Pandya+17}).
Figure~\ref{Fig1:zlmass} shows the scatter plot and the histograms of stellar mass and redshift.

\begin{figure}[H]
   \centering
   \includegraphics[width=12cm, angle=0]{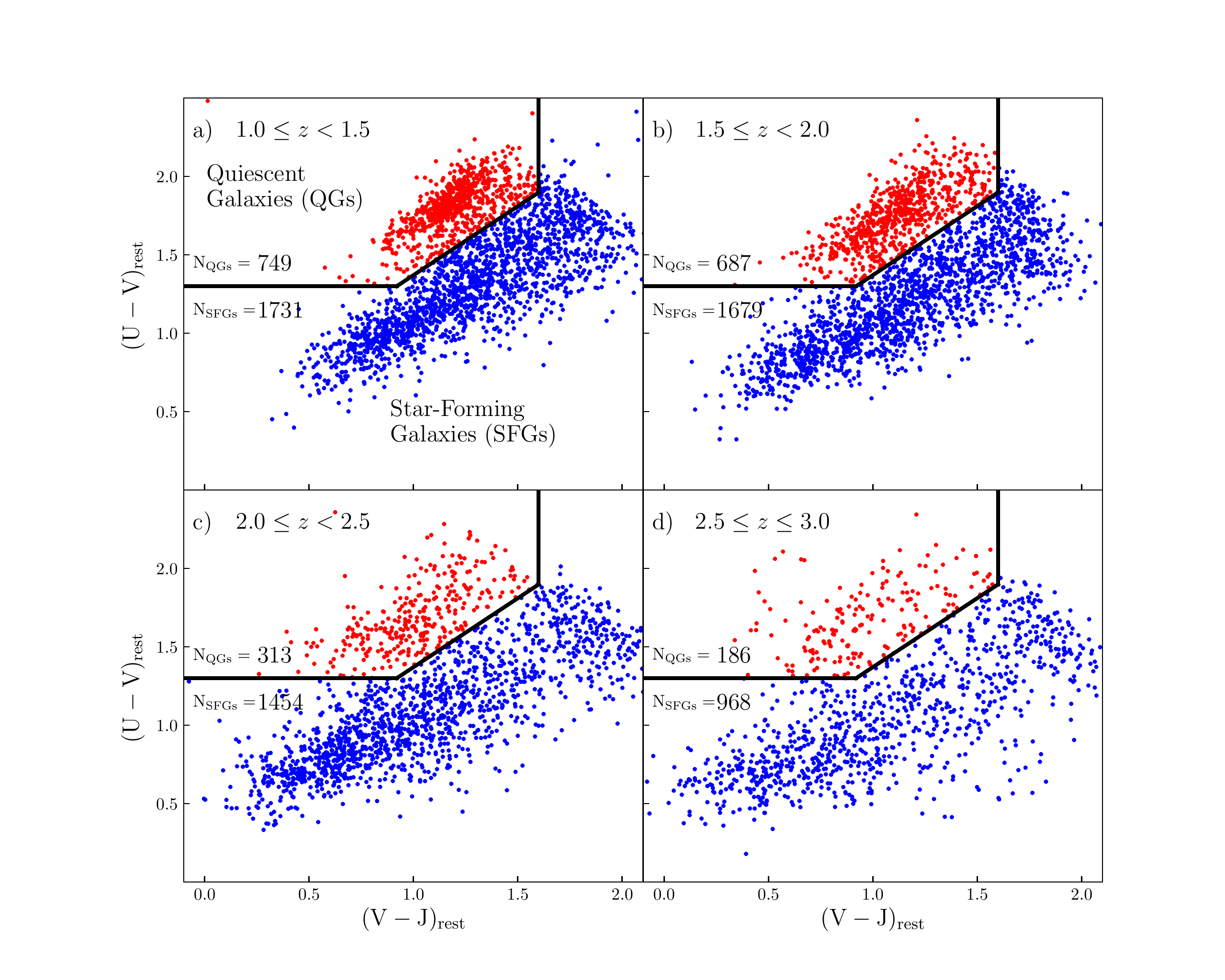}
   \caption{The rest-frame UVJ diagram for massive galaxies in four redshift bins. The solid black lines, following the criteria from \cite{Williams+2009}, separate massive galaxies into quiescent galaxies (red dots) and star-forming ones (blue dots). The numbers of QGs and SFGs are shown near horizontal boundaries.}
   \label{Fig2:UVJ}
   \end{figure}

In order to investigate the evolution of number density of cSFGs and cQGs respectively, we divide our sample into quiescent and star-forming galaxies by using the rest-frame \textrm{UVJ} diagram. Many previous works have suggested that the \textrm{UVJ} diagram can distinguish the QGs from the dusty SFGs, even at higher redshifts $z \sim 3 $ (\citealt{Wuyts+2007,Williams+2009, Brammer+2011, Whitaker+2011,Whitaker+2012,Muzzin+2013,van+der+Wel+2014,Huertas-Company+2015}). The criteria of selecting QGs are shown below (\citealt{Williams+2009}):
\begin{equation}
  ( U - V ) > 0.88 \times ( V - J ) + 0.49,
\label{eq1:select-QG1}
\end{equation}
\begin{equation}
  ( U - V ) > 1.3,
\label{eq2:select-QG2}
\end{equation}
\begin{equation}
  ( V - J ) < 1.6.
\label{eq3:select-QG3}
\end{equation}
In Figure~\ref{Fig2:UVJ}, the rest-frame UVJ diagrams (i.e., $\textrm{U} - \textrm{V}$ vs. $\textrm{V} - \textrm{J}$) are exhibited for four redshift bins with an interval of $\Delta z = 0.5$.
As a result,  5832 SFGs and 1935 QGs with $M_{\ast} \geqslant 10^{10} M_{\odot}$ at $1.0<z<3.0$ are picked up for subsequent selection of compact galaxies.

\subsection{Compactness Criteria}
\label{sect2.3:Compactness critera}
In recent literatures, final results about the abundance of compact galaxies depend heavily on the definition of compactness. There are many versions of compactness criteria which are dramatically different.
One of our objectives is to untangle the effect of different compactness criteria on the abundance of massive compact galaxies.  Various criteria of compact galaxies are addressed in this subsection.

\begin{figure}
   \centering
   \includegraphics[width=14cm, angle=0]{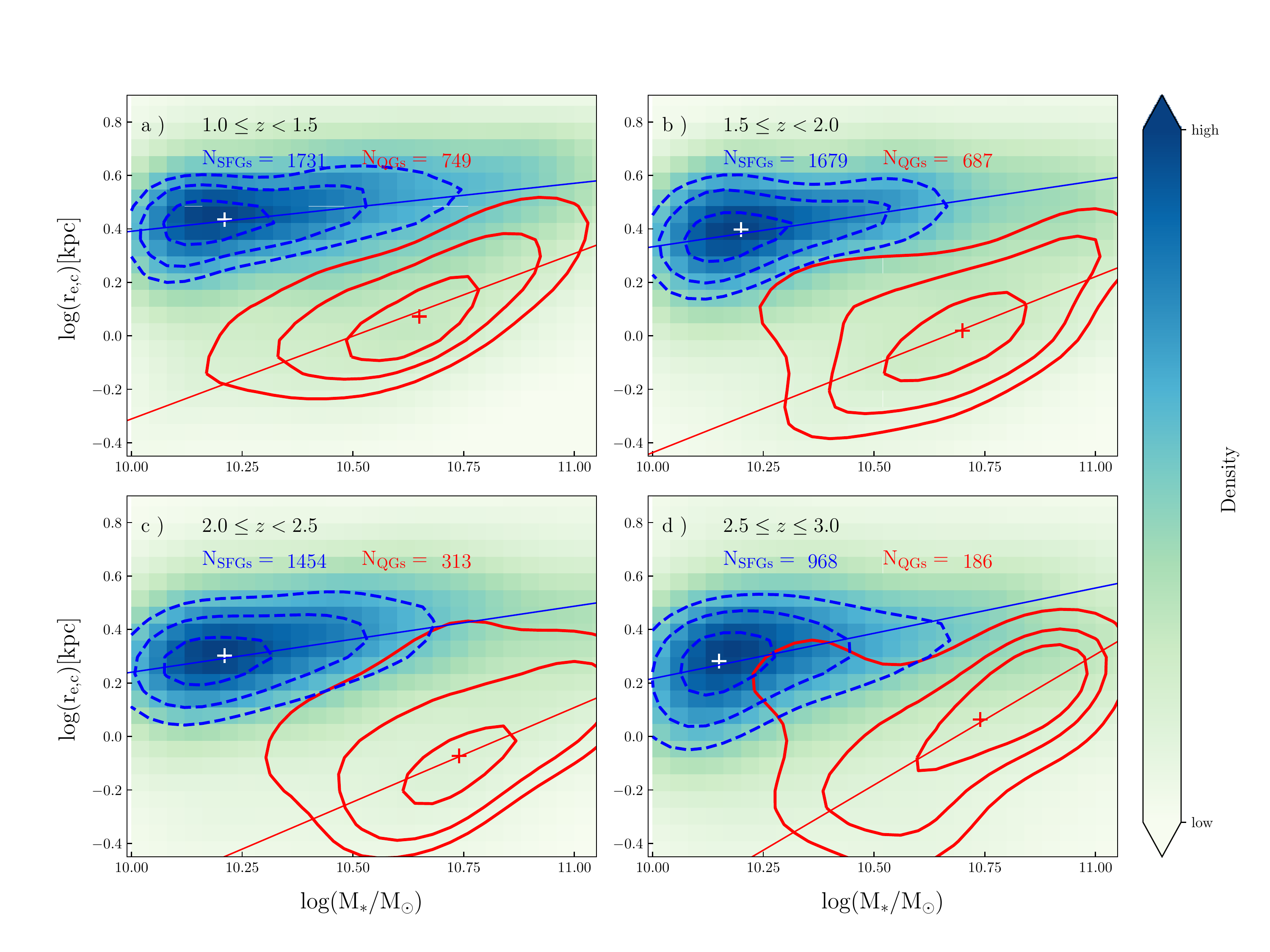}
   \caption{The size of galaxy as a function of stellar mass in four redshift bins. The color blocks represent the relative density in the size-mass relation. The contour levels for SFGs (in blue) and QGs (in red) embrace the $10 \%$, $30 \%$, and $50 \%$ of grid counts with the densities sorted in descending order. Blue and red solid lines show the linear fittings for two inner contours. The median values of circularized effective radius and stellar mass for the samples of SFGs (in white) and QGs (in red) are marked with the crosses. The numbers of SFGs  and QGs are also given in the top.}
   \label{Fig3:contour-z-lmass}
   \end{figure}

It is necessary to describe the structural parameters adopted in the definition of compactness. For quantifying the size of galaxies, the non circularized effective radius $r_{\rm e}$, defined as the semi-major axis in arcsec of the ellipse that contains half of the total light, can be estimated with the fitting with the S\'{e}rsic model (\citealt{van+der+Wel+2012}).
The circularized effective radius, $r_{\rm e,c}$, can be derived by following formula:
\begin{equation}
   r_{\rm e,c} = r_{\rm e} \times \sqrt{q},
\label{eq4:re-re,c}
\end{equation}
where $q$ means the axis ratio, i.e., $q = b/a$. Both $r_{\rm e}$ and $r_{\rm e,c}$ are in units of \textrm{kpc} in this paper.
The size-mass relations for our sample of massive galaxies in four redshift bins are presented in Figure~\ref{Fig3:contour-z-lmass}.  In general, the SFGs have lager sizes than the QGs in all redshift bins. Linear fittings are performed for massive SFGs and QGs, respectively. For both QGs and SFGs, their sizes tend to become larger over cosmic time (i.e., from high to low redshifts). Similar results have been shown in recent works (\citealt{Daddi+2005,Whitaker+2012,Huertas-Company+2015,Gu+2018a,Damjanov+2018inpreparation}).
Compared with the SFGs, massive QGs at $1 < z < 3$ are found to have smaller sizes that are more dependent on the stellar mass. The slopes of the size-mass relation for early-type quiescent population are steeper than those for SFGs, which is in good agreement with \cite{van+der+Wel+2014}.

The \textit{Gini} coefficient, as a nonparametric measuremnt, has been taken to exclude some cSFGs with visually extended structures by \cite{Fang+2015}. As described in \cite{Abraham+2003} and \cite{Lotz+2004}, the \textit{Gini} coeffiicent is defined to quantify the relative distribution of the pixel fluxes:
\begin{equation}
   \textit{Gini} = \frac{\sum^{N}_{l}(2l-N-1)\mid F_{l}\mid}{\overline{F}N(N-1)},
\label{eq5:Gini}
\end{equation}
where $F_{l}$ is the pixel flux value sorted in ascending order, $\overline{F}$ is the mean pixel flux, and $N$ is the total number of pixels belonging to a galaxy.  The \textit{Gini} coefficient can be regarded as a generalized measure of concentration. Moreover, it is able to describe the arbitrary shape of the galaxy without requiring a single well-defined nucleus (i.e, multiple cores). In this work, the \textit{Gini} coefficients are measured by the developed version of Morpheus software (\citealt{Abraham+2007}).

We collect all specific definitions of compact galaxies adopted in recent works. These definitions take different lower limits of stellar mass and different size cuts.
The specific compactness criteria are listed in Table~\ref{Tab1:compactness}, as well as the number counts of our cQG and cSFG samples at $1 < z <3$ for each compactness definition.

\begin{table}[h]
\begin{threeparttable}
\renewcommand\arraystretch{1.5}
\begin{center}
\caption[]{Various Definitions of Compactness and the Sizes of Our cQG and cSFG Samples}\label{Tab1:compactness}
 \begin{tabular}{cccccc}
  \noalign{\smallskip}\hline\hline
No. &   Mass    &  Compactness    & Number  &  Number  &  Abbreviations$^a$\\
   &   limit   & criteria         & of cQGs & of cSFGs &  \\
  \hline\noalign{\smallskip}
1 & $> 10^{10.5}M_{\odot} $     & the ``\textit{most}" compact $r_{\rm e} < 1.4 \rm \; kpc$  & $365$  & $100$ & C13 most \\
2   &  $> 10^{10.5}M_{\odot}$               & the ``\textit{less}" compact $r_{\rm e} < 2.0 \rm \; kpc$  &$587$  & $217$  & C13 less \\
3  & $ > 10^{10.9}M_{\odot}$  & $r_{\rm e, c} < 1.5\;\rm kpc $ &$94$  & $37$  &  QT13\\
4  & $> 10^{10}M_{\odot}$    &   ${\Sigma_{1.5}}^{\rm b} > 10.45 M_{\odot}\;\rm kpc^{1.5}$ & $982$ & $455$ &  B14 \\
5  & $> 10^{10.7}M_{\odot} $ & the ``\textit{most}" compact $r_{\rm e} < 1.5 \times (M_{\ast}/ 10^{11} M_{\odot})^{0.75} $  & $109$  & $47$&  vdW14 most\\
6  & $> 10^{10.7}M_{\odot}$     & the ``\textit{less}" compact $r_{\rm e} < 2.5 \times (M_{\ast}/ 10^{11} M_{\odot})^{0.75} $ &$362$  &$192$ & vdW14 less  \\
7  &   $\geqslant 10^{10}M_{\odot}$   &  ${\Sigma_{1.5}}^{\rm b} \geqslant 10.45 M_{\odot}\;\rm kpc^{1.5}$ and  ${\textit{Gini}} \geqslant 0.4$ & $887$  & $360$   & F15\\
8  &   $> 10^{10.6}M_{\odot} $   &  $\log_{\rm 10}(r_{\rm e, c}) < \log_{\rm 10}(M_{\ast} / M_{\odot}) - 10.7$  & $438$  & $250$& vD15  \\

  \noalign{\smallskip}\hline\hline
\end{tabular}
\begin{tablenotes}
\item[$\rm a$] The different criteria are expressed in abbreviations.( C13: \citealt{Carollo+2013}; QT13: \citealt{Quilis+Trujillo+2013}; B14: \citealt{Barro+2014}; vdW14: \citealt{van+der+Wel+2014}; F15: \citealt{Fang+2015}; vD15: \citealt{van+Dokkum+2015}). The 'most' and 'less' represent the most and less compact criteria as it applies.
  \item[$\rm b$] The $\Sigma_{1.5}$ is a \textit{pseudo}-stellar mass surface density, defined as $\log_{\rm 10} (M_{\ast}/r_{\rm e,c}^{1.5})$ (\citealt{Barro+2013}).
\end{tablenotes}
\end{center}
\end{threeparttable}
\end{table}

\section{The Abundance of massive compact galaxies}
\label{sect:The Abundance of massive compact galaxies}

\subsection{The Fractional Abundance}
\label{The fractional abundance}

To show the effect of compactness definition on the cQG and cSFG abundance at high redshifts, we adopt eight different definitions of compact galaxies  (see Table~\ref{Tab1:compactness}) to select the cQGs and cSFGs at $1 < z < 3$ in the 3D-HST/CANDELS fields.
Fractional abundance of the cQGs (cSFGs) is defined as the ratio of the number of cQGs (cSFGs) to total number of QGs (SFGs).

   \begin{figure}
   \centering
   \includegraphics[width=10cm, angle=0]{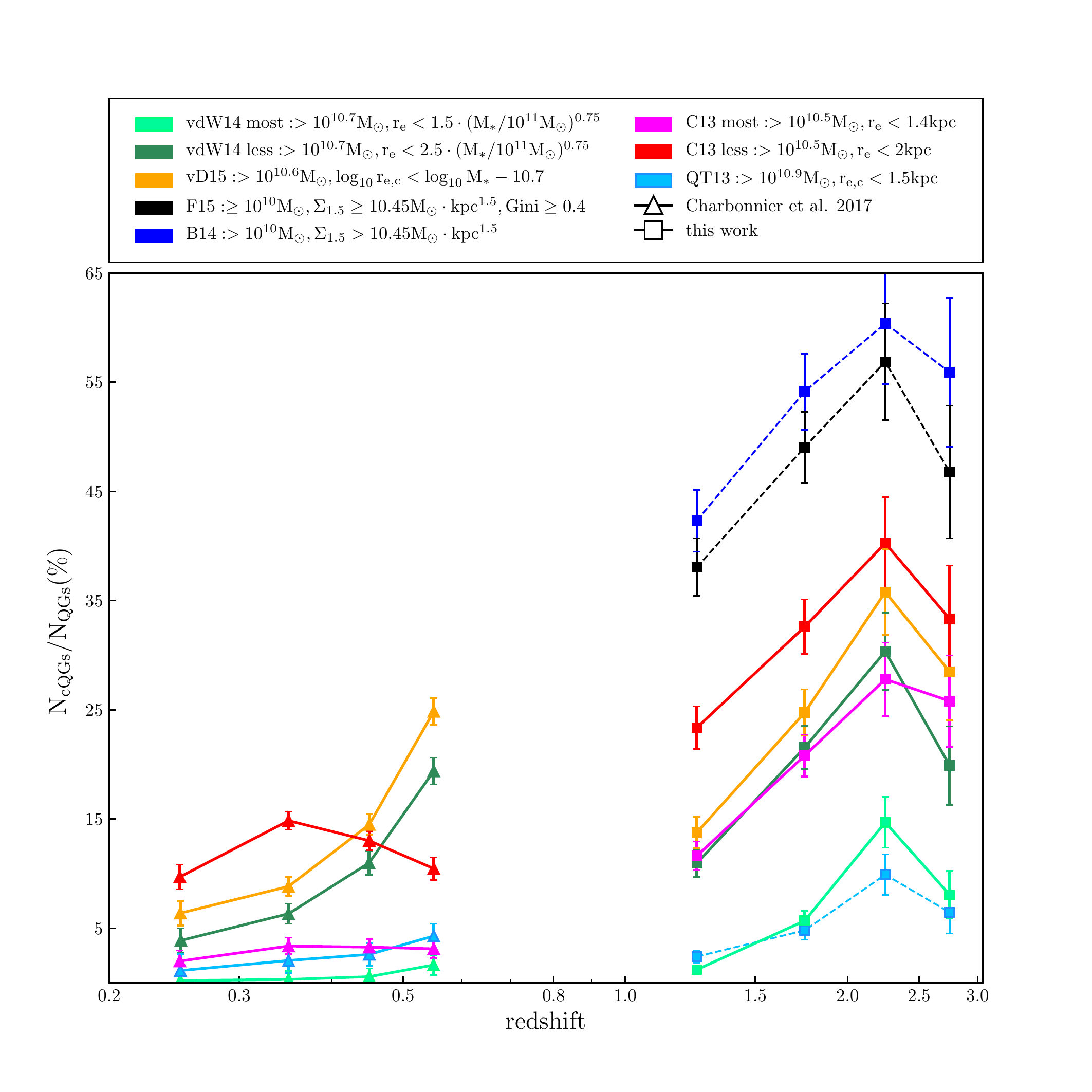}
   \caption{The redshift evolution of the cQG fractional abundance (i.e., the ratio of the number of massive cQGs to that of massive QGs) in a wide range of redshift ($0.2 < z < 3.0$). The abbreviations and specific criteria of the cQGs listed in Table~\ref{Tab1:compactness} are shown in the top of the figure, distinguished by different colors. The data points at high redshifts ($1<z<3$), denoted with the squares, represent the results for our eight samples of cQGs. The results at low redshifts ($0.2<z<0.6$) by \cite{Charbonnier+2017} are also presented with the triangles.}
   \label{Fig4:FcQGs}
   \end{figure}

Compared with \cite{Charbonnier+2017} and \cite{Damjanov+2018inpreparation}, we adopt more criteria to select compact galaxies from massive QGs and SFGs, and the corresponding counts of cQGs and cSFGs are given in Table~\ref{Tab1:compactness}.
Figure~\ref{Fig4:FcQGs} shows the fractional abundances of eight cQG samples at high redshifts which are selected by different compactness criteria.
Except for the compactness definition (QT13: dodgerblue lines) by \cite{Quilis+Trujillo+2013}, it is found that the cQG fractions in the QG samples tend to increase with redshift at $z \gtrsim 2.0$, then decrease rapidly at $z < 2$. Although the different compactness criteria are adopted, overall variations of the cQG fraction with redshift are similar.

In order to observe the cosmic evolution of the cQG fraction from $z \sim 3 $ to 0.2, the fractional abundances at $0.2 <z<0.6$ which were derived by \cite{Charbonnier+2017} are also presented in the same diagram.
Compared with results from above two works, in which the used criteria are from \cite{Carollo+2013} (C13 most: magenta lines and C13 less: red lines), the fractional abundances of the cQGs tend to be fewer from $z \sim$ 1.0 to 0.6.
The cQG fraction seems to increase when the sizes of compact galaxies are related to stellar masses (vdW14 less and most; vD15). By comparing the evolutionary trends between low and high redshifts, the influence by different compactness criteria on the cQG fraction can not be ignored over the blank range of redshift.
Furthermore, for the criteria (QT13: dodgerblue lines) in \cite{Quilis+Trujillo+2013}, the cQG fractions at low and high redshifts  are rather small because of the strict selection of compact galaxies (i.e., the highest mass threshold of $ 10^{10.9}M_{\odot}$ and a small upper limit of circularized effective radius $r_{\rm e,c}=1.5\ \rm kpc$).
It should be mentioned that the diversity of the cQG fraction due to different criteria adopted is found to be larger at $z > 1.0$ (even up to $\sim 50\%$)  than that at $0.2<z<0.6$ (\citealt{Charbonnier+2017}). However, the overall redshift evolution of cQG fraction at $1<z<3$ are similar.

   \begin{figure}
   \centering
   \includegraphics[width=10cm, angle=0]{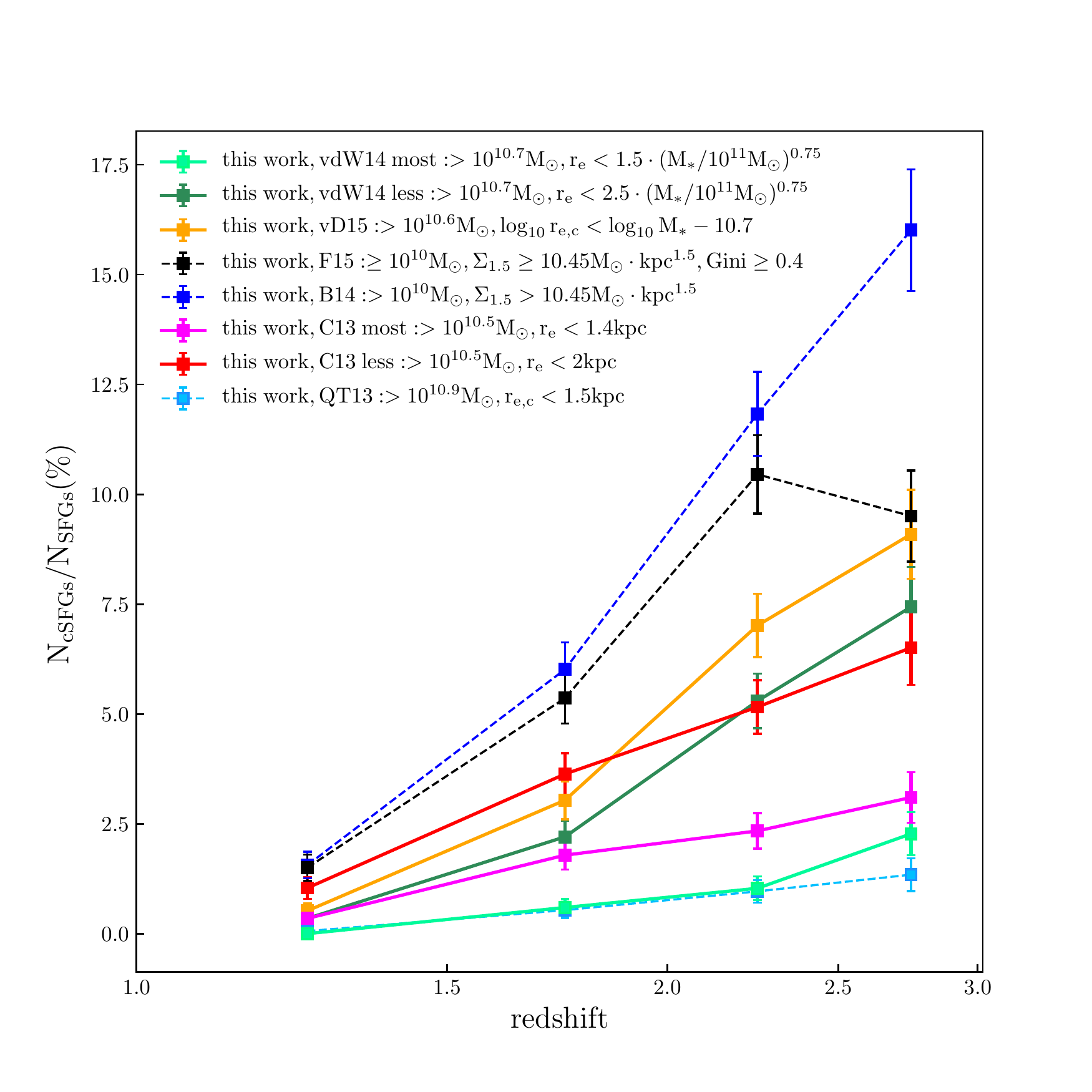}
   \caption{The redshift evolution of fractional abundance of cSFGs between 1.0 and 3.0. The indications of colors and symbols are the same as in Figure~\ref{Fig4:FcQGs}.}
   \label{Fig5:FcSFGs}
   \end{figure}

Compared with cQG fraction, the situation of redshift evolution of cSFG fractions is rather different, as shown in Figure~\ref{Fig5:FcSFGs}. A simple rising trends along redshift can be found for cSFG fractional abundances at $1<z<3$, except for the criterion (F15: black dashed line) from \cite{Fang+2015}.
The trend is broken down at $z>2$ in \cite{Fang+2015} when the \textit{Gini} coefficient is adopted to get rid of some cSFGs with extended structure at high redshifts. The rising slopes in the diagram of fractional abundance vs. redshift are dependent upon the criteria of compactness.

Regardless of difference in compactness definition, the fractional abundance of cSFGs are found to be much higher at high redshifts ($z>2$) than that at lower redshifts ($z \lesssim 1$).
According to some predictions by simulation, cSFGs are formed by gas-rich, dissipational processes, such as cold accretion from the IGM via violent disk instability (\citealt{Dekel+09a,Dekel+09b}), cold mode accretion (\citealt{Birnboim+Dekel+03,Johansson+12}) and major mergers (\citealt{Hopkins+09,Hopkins+10,Wuyts+2010}).
Star formation in the cSFGs is subsequently quenched by some feedbacks such as AGN feedback (\citealt{Barro+2013,Barro+2014,Kocevski+17}) and stellar winds driven by intense starbursts (\citealt{Tremonti+2007, Heckman+2011}). \cite{Kocevski+17} find that $39.2 \%$ of massive cSFGs host an X-ray detected AGN, which is higher than the incidence of AGN in eSFGs, indicating that AGN feedback helps to decline in the number density of cSFGs. Therefore, these feedback mechanisms imprint evidence that extremely rare cSFGs are found at lower redshifts $0.5<z<1$ (\citealt{Trujillo+2009,Taylor+10,Barro+2013,Trujillo+2014}). And compactness can be treated as a very sensitive predictor of passivity among massive galaxies, particularly at higher reshifts (\citealt{Bell+2012,Williams+2014}).

By synthesizing the cosmic evolution of fractional abundances of cQGs and cSFGs, the connection between cQGs and cSFGs can be discussed.
If we adopt a simple evolutionary model (\citealt{Barro+2013}) (see Section \ref{Number Density Evolution} and Figure \ref{Fig8:Transition}), it can be found that the lifetimes of the cSFGs at high redshift selected by different compactness criteria are less than 0.8 Gyr, which are in agreement with \cite{Barro+2013,Barro+2014} and \cite{van+Dokkum+2015}.
Based on the number densities of  green valley galaxies and quiescent galaxies at $0.5<z<2.5$ in the fields of CANDELS, \cite{Gu+2018b} estimated the upper limit of the average transition/quenching time-scale as a function of redshift, and the average quenching time-scale at $z\sim2.5$ is less than 0.35 Gyr.
The fractional abundance for the cQGs peaks at $z \sim 2.0$, which can be construed by the assumption that a certain percentage of the cSFGs at $z \gtrsim 2$ may have been quenched into the cQGs via rapid violent dissipational process (\citealt{Barro+2013,Fang+2015,Williams+2015}).
The average quenching time-scale become longer than 1.3 Gyr since $z \sim 2$, and the accumulative effects from the above-mentioned feedback mechanisms and minor mergers during longer passive evolution may result in a looser stellar distribution in the massive QGs (\citealt{Gu+2018b}).
This picture may help us to understand the declining trend in the cQG fractional abundance since $z\sim 2$.

\subsection{Number Density Evolution}
\label{Number Density Evolution}

   \begin{figure}
   \centering
   \includegraphics[width=10cm, angle=0]{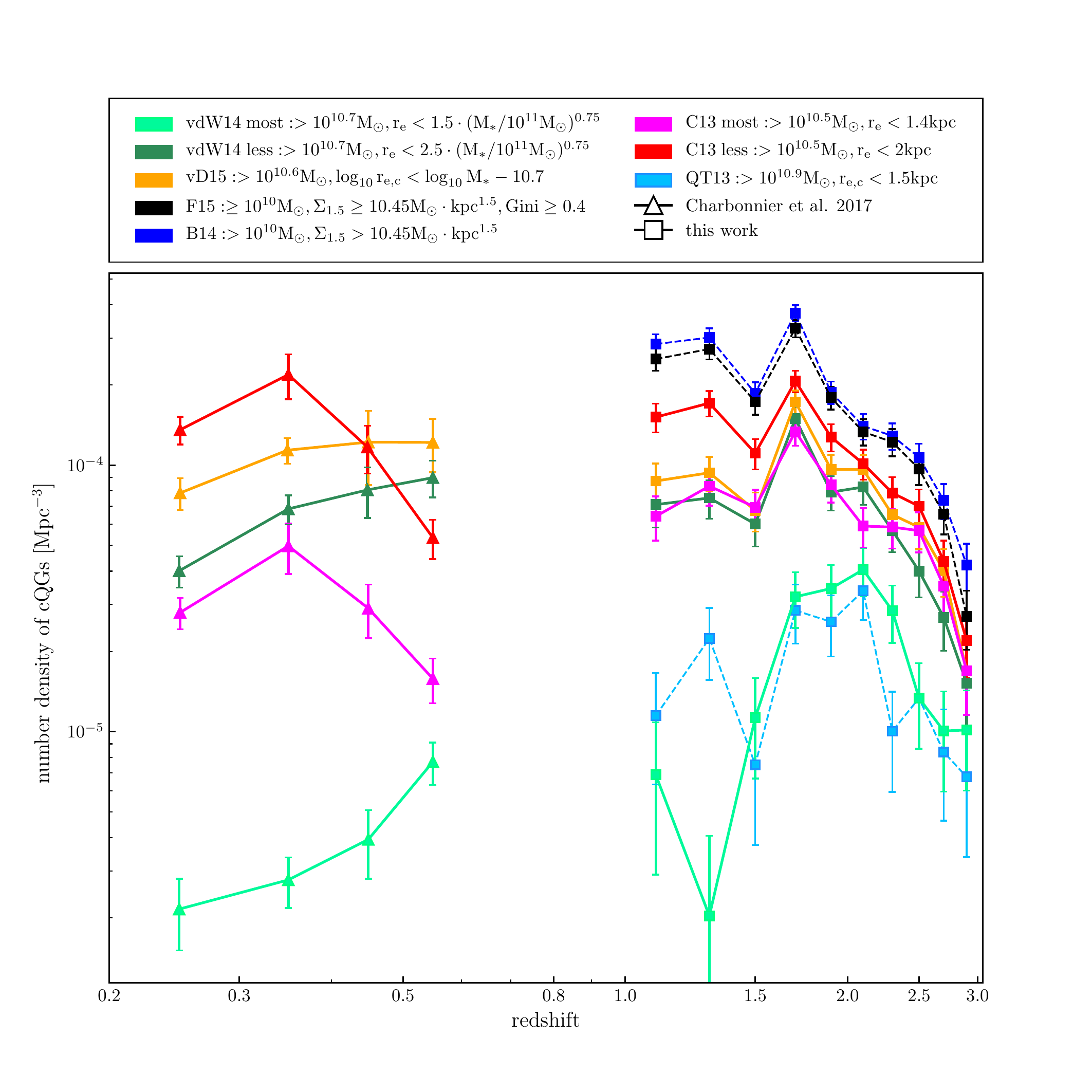}
   \caption{The redshift distribution of number density of cQGs compared our observed cQGs (square) at $z \geqslant 1.0$ with the ones (triangle) compiled from \cite{Charbonnier+2017} at $z < 0.6$. Different colours indicate different definitions of compactness. The indications of colors and symbols are the same as in Figure~\ref{Fig4:FcQGs}.}
   \label{Fig6:Ndensity of cQGs}
   \end{figure}

It has been widely appreciated that massive cSFGs will rapidly quench into the cQGs at high redshifts (\citealt{Barro+2013, Fang+2015, van+Dokkum+2015}). However, the opposite path of evolution that the cQGs begin their star formation activities via accreting new gas has also be proposed (so-called ``rejuvenation'') (\citealt{Graham+2015, Zolotov+2015}). For investigating the evolution of massive compact galaxies at high redshifts, we further quantify the comoving number densities of the cQGs and cSFGs within a small interval of $\Delta z=0.2$.
The number density can be determined by dividing the number of massive compact galaxies by its comoving volume within the redshift interval.
The correction to number density is not adopted in our work due to the high completeness at high redshift (see Section \ref{sect2.2:Sample of Massive QGs and SFGs}). On the opposite, if we follow the method adopted by \cite{Charbonnier+2017}, the real results will be blurred at higher redshit in terms of more obvious disadvantage of double Schechter function at the low mass end (\citealt{Ilbert+13}).

Figure~\ref{Fig6:Ndensity of cQGs} presents the number densities of the cQGs in five 3D-\textsl{HST}/\textrm{CANDELS} fields at $1<z<3$, as well as the results compiled from the \textrm{CS82} data at $0.2 < z < 0.6$ by \cite{Charbonnier+2017}, which are uncorrected by completeness factors. Except for two cQG samples (vdW14 most: springgreen and QT13: dodgerblue lines) selected with a higher mass threshold ($M_{\ast} > 10^{10.7} M_{\odot}$) and a small upper limit of size,  the remaining six samples include at least 300 cQGs, and their number densities are more reliable  statistically. For these large samples of the cQGs at $1<z<3$, their redshift evolutions of the cQG number densities are quite similar, exhibiting a sustaining increase from $z \sim 3$ to 2 and a maximum density at $z \sim 1.8$.  This trend is consistent with the results in previous works (\citealt{Cassata+2011,Cassata+2013,Barro+2013,van+der+Wel+2014,van+Dokkum+2014,van+Dokkum+2015}).
For the cQGs at $1 < z < 2$, the cQG number density tend to be constant, with a typical number density of $ \sim 10^{-4} \rm Mpc^{-3}$.

\begin{figure}
   \centering
   \includegraphics[width=10cm, angle=0]{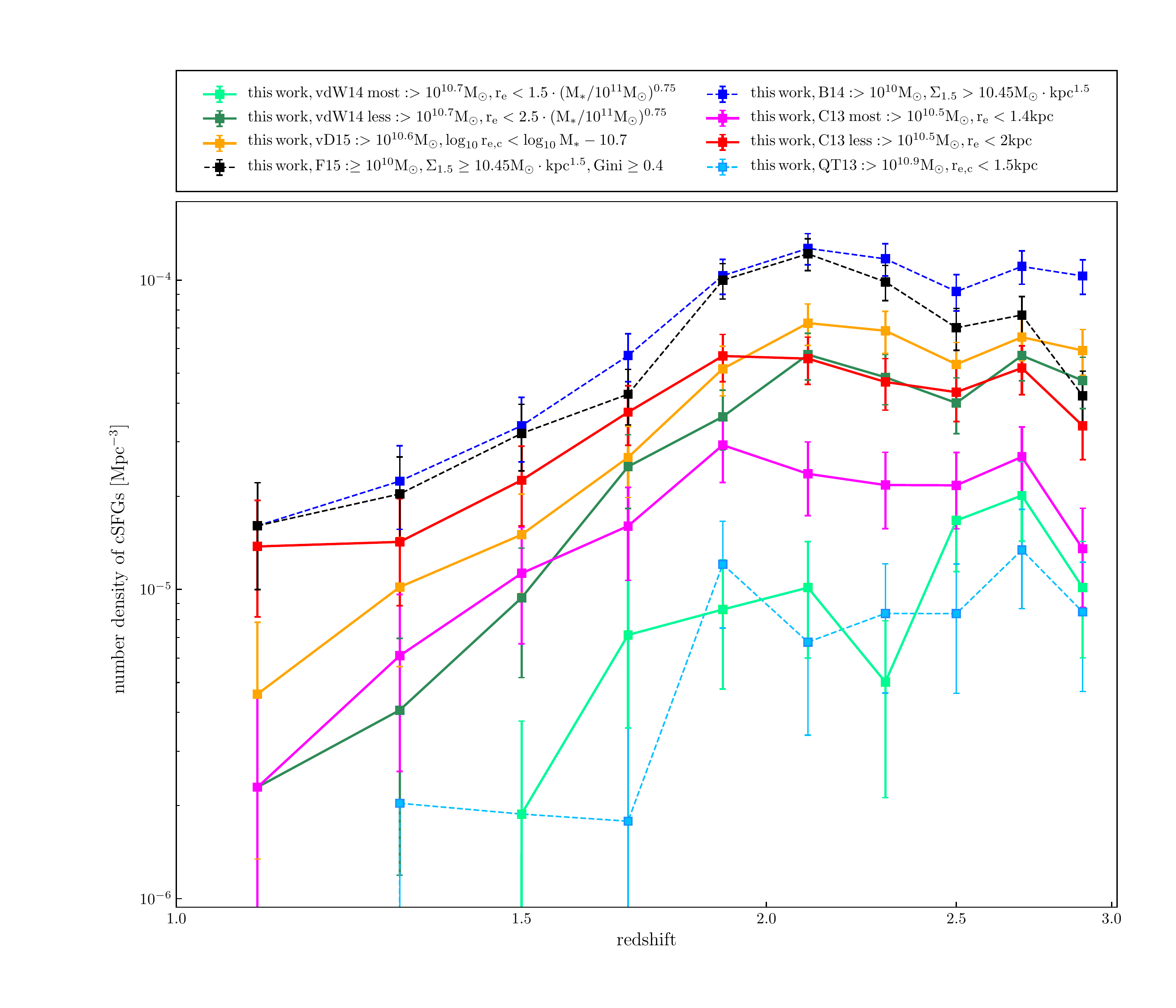}
   \caption{The redshift evolution of number density for the cSFGs at $1<z<3$. All the results are based on our samples in five 3D-\textsl{HST}/CANDELS fields. The indications of colors and symbols are the same as in Figure~\ref{Fig4:FcQGs}.}
   \label{Fig7:Ndensity of cSFGs}
   \end{figure}

Compared to the cQG number densities by \cite{Charbonnier+2017}, we find that the number density of the less compact samples of \cite{van+der+Wel+2014} and \cite{Carollo+2013} are average 0.4 and 0.2 dex higher than their number density under most compactness criterion at $1<z<3$, which are smaller than the deviation values of number density between less and most compact criterion from compiled \cite{Charbonnier+2017}. From the Figure \ref{Fig6:Ndensity of cQGs}, the difference of number density between less and most compact definition (e.g., vdW14 less and most) is obviously getting bigger with decreasing redshift ($1<z<3$), which is likely due to the decrease of number of massive compact galaxies satisfied criteria with higher mass threshold.
The bigger error bars with decreasing redsift may reflect more obvious influence of cosmic variance on lower redshift .
Moreover, if we take a lower mass threshold (i.e., $M_{\ast} \gtrsim 10^{10.5} M_{\odot}$, \citealt{Carollo+2013}), a declining trend over cosmic time within the blank redshift range (i.e., from $z \sim 1.0$ to $0.6$) can be inferred, which is in agreement with \cite{Barro+2013}, \cite{van+der+Wel+2014}, \cite{van+Dokkum+2015} and \cite{Cassata+2013} (for ultra-compact ETGs).
However, for the other cQG definitions with higher mass thresholds (i.e.,$M_{\ast} \gtrsim 10^{10.6} M_{\odot}$, \citealt{van+der+Wel+2014,van+Dokkum+2015}), a constant number densities can be expected at $0.6<z<1.0$, which is consistent with \cite{Cassata+2013} (for compact ETGs) and \cite{Gargiulo+16} (for ultramassive dense ETGs).
Moderate decrease of the cQG number density since $z \sim 1$ can be interpreted with the early-track described in \cite{Barro+2013}, where some cQGs transform into extended quiescent galaxies through minor merger as well mentioned by \cite{Gargiulo+16}.
\cite{Naab+09} have performed hydrodynamic cosmological simulations of the formation of massive galaxies to prove that minior merger may be the main driver for the evolution in sizes and densities of massive early-type galaxies, which is in agreement with \cite{Oser+12} where dry minior mergers come to be predominant since $z \sim 2$ instead major mergers alone.
The rarity of cSFGs since $z \sim 1$ (see Figure~\ref{Fig7:Ndensity of cSFGs}) results in a very low consequent birth rate of the cQGs at lower redshifts.

The redshift evolution of cSFG number density is also presented in the Figure~\ref{Fig7:Ndensity of cSFGs}. Regardless of various cSFG selection criteria, the redshift evolution of the cSFG number density are very similar: keeping a constant number density at $2<z<3$ and a continuous declining from $z \sim 2$ to 1. Our results are consistent with those in \cite{Barro+2013} and \cite{van+Dokkum+2015}.
When taking the definition from \cite{Fang+2015}, a significant drum at $z \sim 2$ can be found in Figure~\ref{Fig7:Ndensity of cSFGs}.
Owing to taking \textit{Gini} coefficient into consideration, a substantial fraction the cSFGs at $z>2$ with clear extended structures may have been excluded by this strict criterion (see Figure \ref{Fig5:FcSFGs} in \citealt{Fang+2015}).


   \begin{figure}
   \centering
   \includegraphics[width=0.98\textwidth]{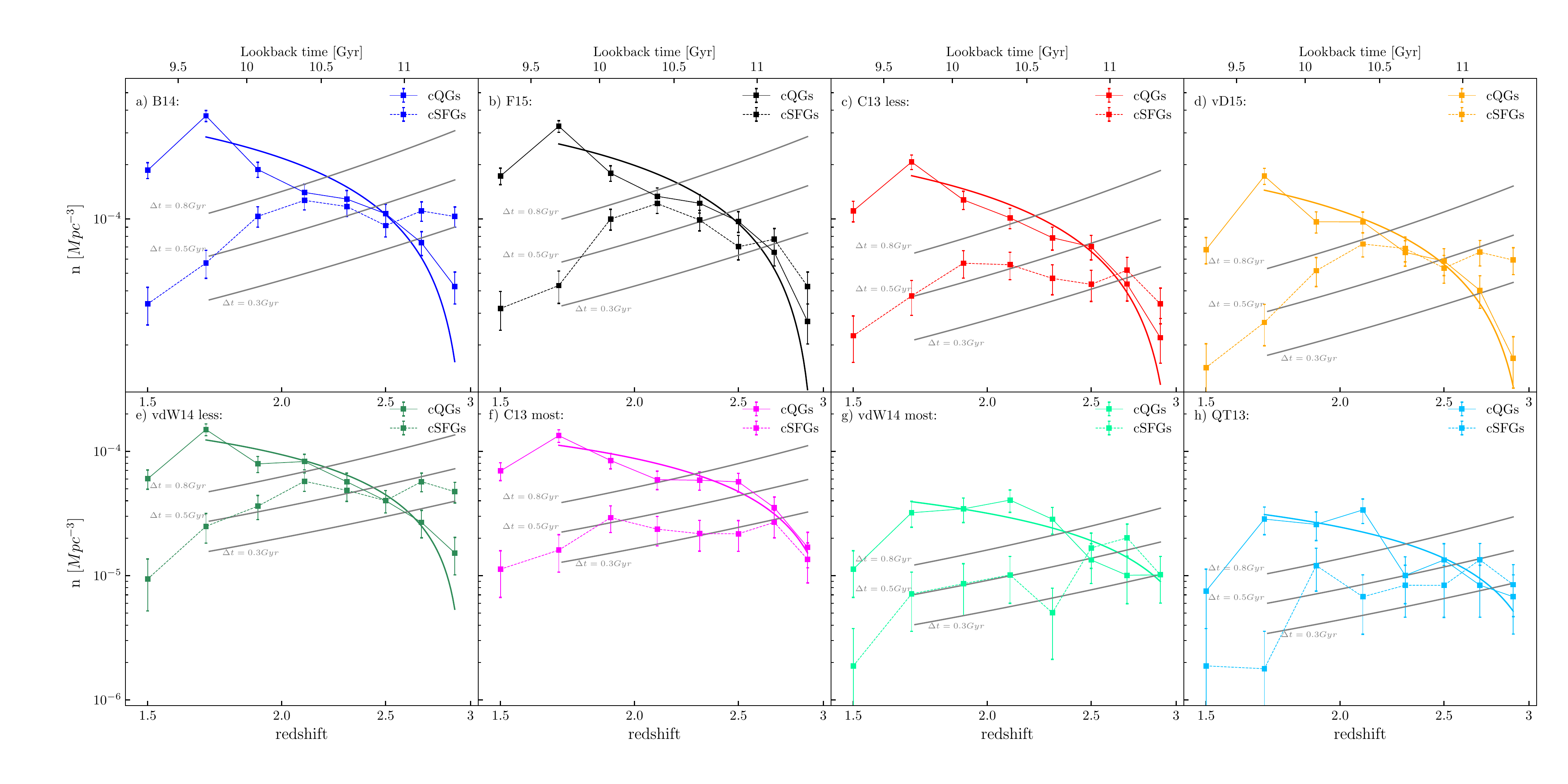}
   \caption{The redshift evolutions of number density for the massive cQGs (solid) and cSFGs (dotted) selected by eight different compactness criteria. Different colours indicate different compactness criteria, and the corresponding abbreviations are shown in the top. The corresponding solid curve is the best fit to the cQG number density in each panel. The solid gray lines depict the evolutions of cSFG number density which are required to match the observed increasing cQG number density, following \cite{Barro+2013}. The corresponding lifetimes of cSFGs and the lookback times (in top row) are shown as well.}
   \label{Fig8:Transition}
   \end{figure}

It's worth considering whether the compact galaxy number density is sensitive to the stellar mass threshold (e.g., QT13: mass limit $>10^{10.9} M_{\odot}$). \cite{Damjanov+15} find that the compactness threshold or the stellar mass range has no significant impact on the compact galaxy number density over the redshift range $0.2<z<0.8$. To check whether the mass limits have an influence on abundance of compact galaxies at higher redshifts ($1<z<3$),  two extreme cases of mass thresholds ($>10^{10.9}$ and $\geqslant 10^{10} M_{\odot}$) are adopted in all compactness definitions.
It is found that all evolutionary trends of compact galaxies (cQGs in Figure \ref{Fig4:FcQGs}, \ref{Fig6:Ndensity of cQGs} and cSFGs in Figure \ref{Fig5:FcSFGs}, \ref{Fig7:Ndensity of cSFGs} ) are not sensitive to the mass thresholds.
Certainly, a higher mass threshold and a strict compactness criterion will lead to few number of compact galaxies selected.

The evolutionary scenario between cQGs and cSFGs can be speculated based on both the redshift evolutions of number density. Figure \ref{Fig8:Transition} presents the difference in the number density evolution between the cQGs and cSFGs at $z \gtrsim 2$. If we adopt a simple evolutionary model proposed by \cite{Barro+2013} that all cSFGs will become quiescent after experiencing a short starburst phase, a crude explanation can be derived for the discrepancy of number density between cQGs and cSFGs.
A plateau in the cSFG number density at $2<z<3$ can be explained by the balance of the birth rate of new cSFGs via rapid gas-rich dissipational process (such as major merger and disk instabilities) and the quenching rate of the cSFGs. The quenching of the cSFGs at $z\gtrsim 2$ will surely lead to a strong increasing in the number density of the cQGs from $z \sim 3$ to 1.7.
As shown in Figure \ref{Fig8:Transition}, it is found that the corresponding lifetimes of cSFGs are approximately $\Delta t_{\rm burst} \sim 0.3-0.8$ Gyr. \cite{Barro+2013} and \cite{van+Dokkum+2015} compared the redshift evolution between the cSFGs and cQGs as well, and estimated the time-scale of quenching via central starburst feedback. They referred an average quenching time-scale below 1 Gyr as the lifetime of cSFGs, which is consistent with our time-scale estimate, ($\Delta t_{\rm burst} \sim 0.3-0.8$ Gyr), for different compactness criteria in this work. By using semianalytic models of galaxy formation, \cite{Barro+2014} suggest that cSFGs would have to end their lives with an abrupt decline of the star formation rate (SFR) on a short timescales ($t_{\rm q} \sim 400$ Myr), and then reproduce the emergence of the quiescent population.
Moreover, for the cSFGs with higher compactness, Figure \ref{Fig8:Transition} shows that the time-scale of current burst of star formation tends to be shorter.
The drop of the cSFG number density since $z \sim 1.8$ will lead to a plateau in the cQG number density at $1.0<z<1.7$ since no new cQGs are added to the sample.
Due to the lower SFR toward decreasing redshift as predicted by some simulations (\citealt{Finlator+07, Tonini+12, Barro+2014}), cSFGs are not formed in large numbers at $z<2$ in the late-track described by \cite{Barro+2013}, and some quiescent galaxies are supplied by the quenching of the eSFGs.
The abundances of cSFGs rapidly drop from $z \sim 2 $ to 1, which can be interpreted by the decrease of gas reservoirs in dark matter halos (\citealt{Croton+2009,Geach+2011}) and the consequent low efficiency of gas-rich dissipation (\citealt{Barro+2013}).
The decline of the abundances of cSFGs in this work corresponds to the decrease of gas-rich major merger rate of massive SFGs since $z\sim 1.8$ derived by \cite{Lopez+13}.

\section{Summary}
\label{sect:summary}
In this paper, a large sample of massive galaxies with $M_{\ast} \geqslant 10^{10} M_{\odot}$ at $1 < z < 3$ in five 3D-\textsl{HST}/CANDELS fields has been separated  into quiescent and star-forming populations by the rest-frame \textrm{UVJ} diagram. We further select the cQGs and cSFGs using eight different definitions of compactness in literatures (\citealt{Carollo+2013,Quilis+Trujillo+2013,Barro+2014,van+der+Wel+2014,Fang+2015,van+Dokkum+2015}). To explore the evolutionary connection between the cQGs and cSFGs, fractional abundance and number density are quantified as a function of redshift. Main conclusions are summarized as below:
\begin{enumerate}
  \item We confirm that massive QGs are on average smaller than massive SFGs in size at $1< z< 3$. for a specified redshift range, the slope of the size-mass relation is steeper for massive QGs. The sizes of massive QGs are much more dependent on stellar mass than those of massive SFGs.
  \item
      We adopt eight different definitions of compact galaxies to select cQGs and cSFGs at $1<z<3$. The effect of compactness definition on the values of fractional abundance and comoving number density is remarkable for the cQG and cSFG samples.  However, except for the compactness definition by \cite{Quilis+Trujillo+2013}, their evolutionary trends in the abundance of compact galaxies are found to be rather similar regardless of the adopted mass thresholds and specific compactness criteria.
  \item
      For compact quiescent galaxies (cQGs), both the fractional abundance and the number density of cQGs peak at $z \sim 2.0$. For the large samples of cQGs, their number densities exhibit a sustaining increase from $z \sim 3$ to 2 and a plateau at $1<z<2$.
      Comparing with the results at $0.2 < z < 0.6$ from \cite{Charbonnier+2017}, a declining tend of number density over cosmic time is expected within the reshift gap (i.e, from $z \sim 1$ to 0.6) for the cQG samples with a lower mass threshold of $10^{10.5} M_{\odot}$. A constant cQG number density at $0.6<z<1$ can be inferred for the more massive cQGs with $M_{\ast} \gtrsim 10^{10.6} M_{\odot}$.
  \item
      For compact star-forming galaxies (cSFGs), a rising trend along redshift is found for fractional abundance at $1<z<3$, except for the compactness criterion with the \textit{Gini} coefficient by \cite{Fang+2015}. A plateau in the number density at $2<z<3$ can be found in the cSFG samples, as well as a continuous declining from $z \sim 2$ to 1.
  \item
      Taking the abundances of both the cSFGs and cQGs at $1<z<3$ into consideration, their behaviors in redshift evolution favor the scenario that a certain fraction of the cSFGs at $z \gtrsim 2$ may have been quenched into the cQGs via rapid violent dissipational processes such as major merger or disk instabilities, which leads to a remarkable increasing in the cQG number density from $z \sim 3$ to 2.
      Rarity of the cSFGs at lower redshifts ($z < 1$) is due to the decrease of available gas in dark matter halos. A small fractional abundance for local cQGs ($z<0.3$) may be due to the effect of size enlargement via minor mergers.
\end{enumerate}

\begin{acknowledgements}
This work is based on observations taken by the 3D-HST Treasury Program (GO 12177 and 12328) with the NASA/ESA HST, which is operated by the Association of Universities for Research in Astronomy, Inc., under NASA contract NAS5-26555.
This work is supported by the National Natural Science Foundation of China (Nos. 11673004, 11873032, 11433005) and by the Research Fund for the Doctoral Program of Higher Education of China (No. 20133207110006).
\end{acknowledgements}

\label{lastpage}

\end{document}